\documentclass[12pt]{article}
\usepackage{graphicx}
\usepackage{endfloat}

\begin{document}
\title{ NMR implementation of a quantum scheduling algorithm}
\author{\small{} Jingfu Zhang,$^{1}$ Zhiheng Lu,$^{1}$
Zhiwei Deng,$^{2}$ Hui Wang,$^{2}$
  \\
\small{} $^{1}$Department of Physics,\\
\small{}Beijing Normal University, Beijing,
100875, People's Republic of China\\
\small{} $^{2}$Testing and Analytical Center,\\
\small{}  Beijing Normal University,
 Beijing, 100875, People's Republic of China}
\date{}
\maketitle
\begin{center}
\begin{minipage}{120mm}
\hspace{0.3cm} {\small
  The improved quantum scheduling algorithm
proposed by Grover has been generalized using the generalized
quantum search algorithm, in which a unitary operator replaces the
Walsh-Hadamard transform,  and $\pi/2$ phase rotations replace the
selective inversions, in order to make the quantum scheduling
algorithm suitable for more cases. Our scheme is realized on a
nuclear magnetic resonance (NMR) quantum computer. Experimental
results show a good agreement between theory and experiment.}

 PACS number(s):03.67
\end{minipage}
\end{center}
\vspace{0.3cm}
\section{Introduction}
  The scheduling algorithm solves the intersection problem, i.e.,
the problem is to find common elements in two sets. Let Alice and
Bob be two distant parties who wish to collaborate on a common
task. They each have a schedule listing $N$ slots of time. Their
schedules can be represented as two strings if Alice and Bob each
have $N$ classical or quantum bits. For each person, a bit in
state $1$ denotes a slot that is available, and a qubit in state
$0$ denotes a slot that is not available. The intersection problem
becomes to find the common 1s in the two strings. Alice and Bob
need to exchange information to find the common 1s. The problem is
how to reduce the exchanging information
\cite{Brassard,Peter,Klauck}.

  If Alice and Bob exchange classical bits, they will need to
exchange $O(N)$ bits. H. Buhrman et al found that Alice and Bob
could find the common 1s by exchanging $O(\sqrt{N}\log_{2}N)$
qubits, if they exploited the quantum parallelism
\cite{BCW,Grover97}. Grover proposed an improved algorithm in
theory for the special case in which the string of either Alice or
Bob has few 1s \cite{Grover02}. Grover's theoretical scheme can be
described as follows. Assume that there is a $\log_{2}N$ qubit
register, by setting $N=2^{n}$, where $n$ is an integer. The
register can be transmitted between Alice and Bob. Assume that
Alice has $\epsilon N$ 1s in her $N$ bit string, where $\epsilon
N\ll N$. For convenience, we assume that Bob also has $\epsilon N$
1s in his $N$ bit string, and there is a single common 1 in the
two strings. Alice encodes the $N$ slots in the register, by
applying the Walsh-Hadamard  transform (denoted as W) to
$|\overline{0}>$, where $|\overline{0}>$ denotes that all qubits
in the register lie in state $|0>$. After she repeats
$WI_{\overline{0}}WI_{A}$ $m$ times, where $m=
\pi\sqrt{1/\epsilon}/4$ \cite{Boyer}, the register lies in the
superposition corresponding to Alice's available slots. $I_{A}$
denotes the selective inversion (or suitable inversion) for the
basis states corresponding to Alice's available slots.
$I_{\overline{0}}$ denotes the selective inversion for
$|\overline{0}>$. The composite transformation used by Alice is
denoted by $G$, represented by $G\equiv
(WI_{\overline{0}}WI_{A})^{m}W$. According to the generalized
quantum search algorithm  \cite{Grover98},
$(GI_{\overline{0}}G^{-1}I_{B})^{m}G$ transforms $|\overline{0}>$
to the state corresponding to the available slot of the both of
Alice and Bob, where $G^{-1}$ denotes the inversion of $G$, and
$I_{B}$ denotes the selective inversion for the basis states
corresponding to Bob's available slots.

  Alice can carry out $W$, $I_{A}$, and $I_{\overline{0}}$. When she needs
$I_{B}$, she must send the register to Bob, who applies $I_{B}$ to
the register, and then returns it to Alice. One should notes that
the overall state of the register is unaltered during the course
of sending \cite{Klauck}. Obviously, the number of times the
register needs to be sent to Bob is equal to the number of $I_{B}$
operations.

  In the improved scheduling algorithm proposed by Grover, he used
the original quantum search algorithm, which is not efficient or
invalid at all in some cases \cite{Zhangajp}. The limits on the
search algorithm impose restrictions on the quantum scheduling
algorithm. In this paper, we generalize Grover's improved
algorithm by replacing $W$ by a proper unitary operator $U$, and
replacing the selective inversions by $\pi/2$ phase rotations, in
order to make the quantum scheduling algorithm suitable for more
cases. Correspondingly, $I_{\overline{0}}$ is also replaced by a
$\pi/2$ phase rotation for $|\overline{0}>$. We implement our
scheme using nuclear magnetic resonance (NMR).

\section{Generalizing the quantum scheduling algorithm}

  Our experiments use a sample of carbon-13 labelled chloroform
dissolved in d6-acetone. Data are taken at room temperature with a
Bruker DRX 500 MHz spectrometer. The resonance frequencies
$\nu_{1}=125.76$ MHz for $^{13}C$, and $\nu_{2}=500.13$ MHz for
$^{1}H$. The coupling constant $J$ is measured to be 215 Hz. The
Hamiltonian of this system is represented as \cite{Ernst}

\begin{equation}\label{1}
  H=-2\pi\nu_{1}I_{z}^{1}-2\pi\nu_{2}I_{z}^{2}+2\pi J I_{z}^{1}
  I_{z}^{2},
\end{equation}
by setting $\hbar=1$, where $I_{z}^{k}(k=1,2)$ are the matrices
for $z$-component of the angular momentum of the spins. The
evolution caused by a radio-frequency (rf) pulse on resonance
along $x$ or $-y$ axis is represented as
$R_{x}^{k}(\varphi)=e^{i\varphi I_{x}^{k}}$ or
$R_{y}^{k}(-\varphi)= e^{-i\varphi I_{y}^{k}}$, with $k$
specifying the affected spin. The pulse used above is denoted by
$[\varphi]_{x}^{k}$ or $[-\varphi]_{y}^{k}$. The coupled-spin
evolution is denoted as

\begin{equation}\label{2}
  [\tau]=e^{-i2\pi J\tau I_{z}^{1} I_{z}^{2}},
\end{equation}
where $\tau$ is evolution time. The pseudo-pure state

\begin{equation}\label{3}
  |00>=\left(\begin{array}{c}
    1 \\
    0 \\
    0 \\
    0 \
  \end{array}\right)
\end{equation}
is prepared by using spatial averaging \cite{Cory}, where $|0>$
denotes up spin state. The basis states are arrayed as
$|00>,|01>,|10>$, $ |11>$. In experiments, we exploit rf and
gradient pulse sequence to transform the system from the
equilibrium state to the state
$\rho_{0}=I_{z}^{1}/2+I_{z}^{2}/2+I_{z}^{1}I_{z}^{2}$, where
$\rho_{0}$ denotes the deviation density matrix equivalent to
$|00>$ \cite{Zhang03}.

  The Walsh-Hadamard transform is replaced by $U=R_{y}^{1}(\pi/2)R_{y}^{2}(-\pi/2)$,
which is represented as

\begin{equation}\label{4}
  U=\frac{1}{2}\left(\begin{array}{cccc}
    1 & -1 & 1 &-1 \\
    1 &  1 & 1 & 1 \\
    -1 & 1 & 1 & -1 \\
    -1 & -1 & 1 & 1 \
  \end{array}\right).
 \end{equation}
Alice and Bob each have a four bit string, in which the positions
of slots are encoded by a two qubit register. The basis states
$|00>$, $|01>$, $|10>$, $|11>$ correspond to the first, second,
third, and fourth slots, respectively.  Assume that Alice's
available slots are the first and the second ones, which
correspond to $|00>$, and $|01>$, respectively. $I_{A}$ is
replaced by a $\pi/2$ phase rotation \cite{Biham}, represented as

\begin{equation}\label{5}
  I_{A12}^{\frac{\pi}{2}}=\left(\begin{array}{cccc}
    i & 0 & 0 & 0 \\
    0& i & 0 & 0 \\
    0 & 0 & 1 & 0 \\
    0 & 0 & 0 & 1 \
  \end{array}\right).
 \end{equation}
Considering the phase matching condition \cite{Long},
$I_{\overline{0}}$ is replaced by

\begin{equation}\label{6}
  I_{\overline{0}}^{\frac{\pi}{2}}=\left(\begin{array}{cccc}
    i & 0 & 0 & 0 \\
    0& 1 & 0 & 0 \\
    0 & 0 & 1 & 0 \\
    0 & 0 & 0 & 1 \
  \end{array}\right).
 \end{equation}

We assume that there are three cases for Bob's available slots: 1)
The first and the fourth slots are available; 2) The second and
the third slots are available; and 3) The first and the second
slots are available. For the three cases, $I_{B}$ is replaced by

\begin{equation}\label{7}
  I_{B14}^{\frac{\pi}{2}}=\left(\begin{array}{cccc}
    i & 0 & 0 & 0 \\
    0& 1  & 0 & 0 \\
    0 & 0 & 1 & 0 \\
    0 & 0 & 0 & i\
  \end{array}\right),
 \end{equation}

\begin{equation}\label{8}
  I_{B23}^{\frac{\pi}{2}}=\left(\begin{array}{cccc}
    1& 0 & 0 & 0 \\
    0& i  & 0 & 0 \\
    0 & 0 & i & 0 \\
    0 & 0 & 0 & 1\
  \end{array}\right),
 \end{equation}

\begin{equation}\label{9}
  I_{B12}^{\frac{\pi}{2}}=\left(\begin{array}{cccc}
    i & 0 & 0 & 0 \\
    0& i  & 0 & 0 \\
    0 & 0 & 1 & 0 \\
    0 & 0 & 0 & 1\
  \end{array}\right),
 \end{equation}
respectively. By defining the composite operators $G_{12}\equiv
-UI_{\overline{0}}^{\frac{\pi}{2}}U^{-1}I_{A12}^{\frac{\pi}{2}}U$,
$Q_{1}\equiv
-G_{12}I_{\overline{0}}^{\frac{\pi}{2}}G_{12}^{-1}I_{B14}^{\frac{\pi}{2}}G_{12}$,
$Q_{2}\equiv-G_{12}I_{\overline{0}}^{\frac{\pi}{2}}G_{12}^{-1}I_{B23}^{\frac{\pi}{2}}G_{12}$,
and $Q_{12}\equiv
-G_{12}I_{\overline{0}}^{\frac{\pi}{2}}G_{12}^{-1}I_{B12}^{\frac{\pi}{2}}G_{12}$,
we obtain

\begin{equation}\label{10}
G_{12}|00>=e^{-i\pi/4}(|00>+|01>)/\sqrt{2},
\end{equation}

\begin{equation}\label{11}
Q_{1}|00>=-i|00>,
\end{equation}

\begin{equation}\label{12}
Q_{2}|00>=-i|01>,
\end{equation}

\begin{equation}\label{13}
Q_{12}|00> =e^{-i\pi/4}(|00>+|01>)/\sqrt{2}.
\end{equation}
Eq.(\ref{11}), for example, shows that the register is transmitted
between Alice and Bob only one time, and the common slot, which
corresponds to $|00>$, is obtained. Similarly, if Alice's
available slots are the third and fourth ones, we replace
$I_{A12}^{\frac{\pi}{2}}$ by

\begin{equation}\label{14}
  I_{A34}^{\frac{\pi}{2}}=\left(\begin{array}{cccc}
    1 & 0 & 0 & 0 \\
    0& 1 & 0 & 0 \\
    0 & 0 & i & 0 \\
    0 & 0 & 0 & i \
  \end{array}\right)
 \end{equation}
and obtain

\begin{equation}\label{15}
G_{34}|00>=-e^{-i\pi/4}(|10>+|11>)/\sqrt{2},
\end{equation}
where
$G_{34}\equiv-UI_{\overline{0}}^{\frac{\pi}{2}}U^{-1}I_{A34}^{\frac{\pi}{2}}U$.
We also assume that there are three cases for Bob: 1) The first
and the fourth slots are available; 2) The second and the third
slots are available; and 3) The third and the fourth slots are
available. By defining $Q_{4}\equiv
-G_{34}I_{\overline{0}}^{\frac{\pi}{2}}G_{34}^{-1}I_{B14}^{\frac{\pi}{2}}G_{34}$,
$Q_{3}\equiv-G_{34}I_{\overline{0}}^{\frac{\pi}{2}}G_{34}^{-1}I_{B23}^{\frac{\pi}{2}}G_{34}$,
and $Q_{34}\equiv
-G_{34}I_{\overline{0}}^{\frac{\pi}{2}}G_{34}^{-1}I_{B34}^{\frac{\pi}{2}}G_{34}$,
we obtain

\begin{equation}\label{16}
Q_{4}|00>=i|11>,
\end{equation}

\begin{equation}\label{17}
Q_{3}|00>=i|10>,
\end{equation}

\begin{equation}\label{18}
Q_{34}|00> =-e^{-i\pi/4}(|10>+|11>)/\sqrt{2},
\end{equation}
where

\begin{equation}\label{19}
  I_{B34}^{\frac{\pi}{2}}=\left(\begin{array}{cccc}
    1 & 0 & 0 & 0 \\
    0& 1 & 0 & 0 \\
    0 & 0 & i & 0 \\
    0 & 0 & 0 & i \
  \end{array}\right).
\end{equation}
The overall phases before wave functions can be ignored.

\section{Experimental procedure and results}
   The coupled-spin evolution described as Eq.(\ref{2}) is realized by
pulse sequence $\tau/2-[\pi]_{x}^{1,2}-\tau/2-[-\pi]_{x}^{1,2}$
\cite{Linden}, where $[\pi]_{x}^{1,2}$ denotes a nonselective
pulse (hard pulse), and the symbol $\tau/2$ denotes the evolution
caused by the magnetic field for $\tau/2$ when pulses are switched
off. $[\pi]_{x}^{1,2}$ pulses are applied in pairs each of which
take opposite phases in order to reduce the error accumulation
causes by imperfect calibration of $\pi$-pulses \cite{Zhu}. $U$ is
realized by $[\pi/2]_{y}^{1}-[-\pi/2]_{y}^{2}$.
$I_{A12}^{\frac{\pi}{2}}=R_{y}^{1}(\pi/2)R_{x}^{1}(\pi/2)R_{y}^{1}(-\pi/2)$
(up to an irrelevant overall phase factor), and is realized by
$[-\pi/2]_{y}^{1}-[\pi/2]_{x}^{1}-[\pi/2]_{y}^{1}$, noting that
the pulses are applied from left to right \cite{Vandersypen}.
$I_{B14}^{\frac{\pi}{2}}=[7/2J]$,
$I_{B23}^{\frac{\pi}{2}}=[1/2J]$, and
$I_{B12}^{\frac{\pi}{2}}=I_{A12}^{\frac{\pi}{2}}$. By modifying
the pulse sequences in Ref. \cite{Chuangprl, Long}, we find
$I_{\overline{0}}^{\frac{\pi}{2}}=
R_{y}^{1,2}(\pi/2)R_{x}^{1,2}(\pi/4)R_{y}^{1,2}(-\pi/2)[15/4J]$.
By optimizing the pulse sequences, we obtain

$G_{12}=-R_{y}^{1}(\pi/2)R_{y}^{1}(\pi/2)R_{x}^{1,2}(\pi/4)R_{y}^{1,2}(-\pi/2)
         [15/4J]R_{x}^{1}(\pi/2)$,
and

$G_{12}^{-1}=-R_{x}^{1}(-\pi/2)[1/4J]R_{y}^{1,2}(\pi/2)R_{x}^{1,2}(-\pi/4)
               R_{y}^{1}(-\pi/2)R_{y}^{1}(-\pi/2)$.

  Similarly,
$I_{A34}^{\frac{\pi}{2}}=R_{y}^{1}(\pi/2)R_{x}^{1}(-\pi/2)R_{y}^{1}(-\pi/2)$,
and $I_{B34}^{\frac{\pi}{2}}=I_{A34}^{\frac{\pi}{2}}$. $G_{34}$
can be optimized as
$G_{34}=-R_{y}^{1}(\pi/2)R_{y}^{1}(\pi/2)R_{x}^{1,2}(\pi/4)R_{y}^{1,2}(-\pi/2)
         [15/4J]R_{x}^{1}(-\pi/2)$.

   The experimental results are represented by
the deviation density matrices reconstructed from the spectra
recorded through the readout pulses by the technique of state
tomography \cite{Chuang98}. We first prepare pseudo- pure state
$|00>$. The generalized quantum scheduling algorithm starts with
$|00>$. $Q_{1}$, $Q_{2}$, and $Q_{12}$ transform $|00>$ into
$|00>$, $|01>$, and $(|00>+|01>)/\sqrt{2}$, respectively.
Figs.1(a-c) show the experimentally measured deviation density
matrices after $Q_{1}$, $Q_{2}$, and $Q_{12}$ are applied to
$|00>$, respectively. Figs.1(d-f) show the theoretically expected
matrices $|00><00|$, $|01><01|$, and $(|00>+|01>)(<00|+<01|)/2$,
respectively. In contrast with the theoretical expectation, for
the values of the theoretical non-zero elements, the relative
errors of the experimental values in Figs.1 (a) and (b) are less
than $12\%$, and the relative errors in Fig.1 (c) are less than
$22\%$. The relative errors increase in Fig.1(c) because the
theoretical values are only half of those in Figs.1 (a) and (b).
The other small elements in Figs.1(a-c) are less than $30\%$. The
errors mainly result from the imperfection of pulses, effect of
decoherence and inhomogeneity of magnetic field.

\section{Conclusion}
   We have realized the generalized quantum scheduling algorithm on
a two qubit NMR quantum computer, and obtained nontrivial results.
For each case discussed above, Alice and Bob each have two
available slots in their schedules either of which consists of
four slots. The original scheduling algorithm does not work in
this case, because the original search algorithm is invalid in the
case that the number of the marked states is equal to the number
of unmarked states. However, using the generalized scheduling
algorithm, Alice and Bob can find the common slots by exchanging
the register only one time.

\section {Acknowledgements}
 This work was supported by the National Nature Science
Foundation of China. We are also grateful to Professor Shouyong
Pei of Beijing Normal University for his helpful discussions on
the principle of quantum algorithm and also to Dr. Jiangfeng Du of
University of Science and Technology of China for his helpful
discussions on experiment.

\newpage

\newpage
{\begin{center}\large{Figure Captions}\end{center}
\begin{enumerate}

\item Experimentally measured deviation density matrices shown as
Figs.1(a)-(c) and theoretically expected density matrices shown as
Figs.1(d)-(f) after the completion of the quantum scheduling
algorithm, corresponding to operations $Q_{1}$, $Q_{2}$, and
$Q_{12}$, respectively. Only the real component is plotted. The
imaginary portion, which is theoretically zero, is found to
contribute less than $15\%$ to the experimental results.

\end{enumerate}
\begin{figure}{1}
\includegraphics[width=6in]{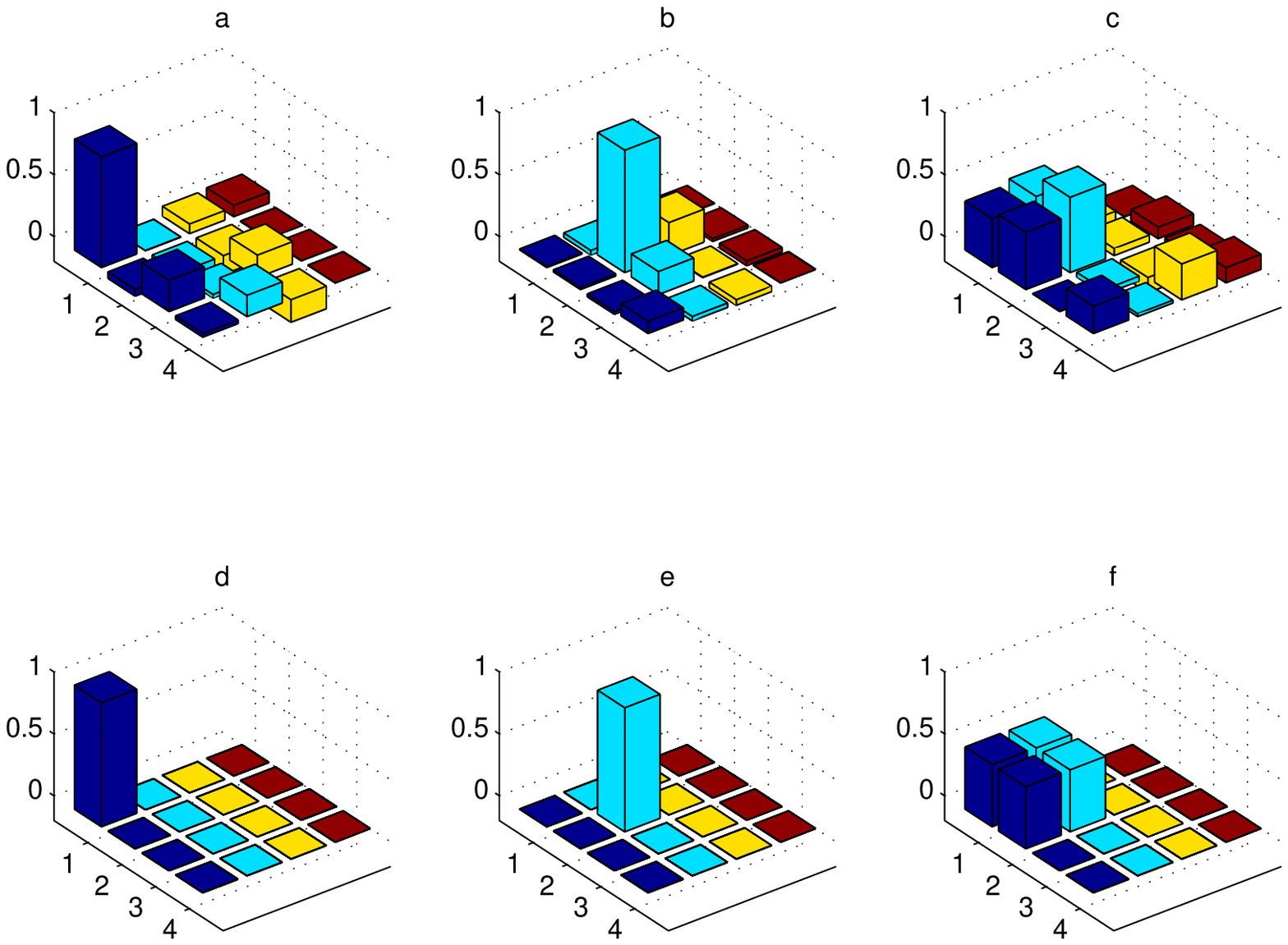}
\caption{\label{1}}
\end{figure}
\end{document}